\documentclass[iop,apj]{emulateapj}

\usepackage{graphicx}         % standard LaTeX graphics tool
\newcommand \kms{km~$\rm{s}^{-1}$}
\newcommand \cc{$\rm{cm}^{-3}$}
\newcommand \mum{$\mu$m}
\slugcomment{Submitted 2011 December 5; accepted  2012 March 2}

\begin{document}
  
 %%%%
%%%%\begin{document}

\title{Spitzer observations of the HH~1/2 system. The discovery of the counterjet}

\author{Noriega-Crespo, A.\altaffilmark{1}}
\author{Raga, A. C. \altaffilmark{2}}

\altaffiltext{1}{Infrared Processing and Analysis Center,
California Institute of  Technology, CA 91125  USA}
\altaffiltext{2}{Instituto de Ciencias Nucleares, Universidad Nacional 
Aut\'onoma de M\'exico, Ap. 70-543, 04510 D.F., M\'exico}

\begin{abstract}
We present unpublished {\it Spitzer} IRAC observations of the HH~1/2 young
stellar outflow processed with a high angular resolution deconvolution
algorithm , that produces sub-arcsecond ($\sim$ 0.6 - 0.8\arcsec) images.
In the resulting mid-infrared images, the optically invisible counterjet 
is detected for the first time. The counterjet is approximately half as bright 
as the jet at 4.5\mum~(the IRAC band that best traces young stellar outflows) 
and has a length of $\sim 10''$. The NW  optical jet itself can be 
followed back in the mid-IR to the position of the exciting VLA~1 source.
An analysis of the IRAC colors indicates that the jet/counterjet emission is 
dominated by collisionally excited H$_2$ pure rotational lines 
arising from a medium with a neutral Hydrogen gas density of
$\sim 1000-2000$~cm$^{-3}$ and a temperature $\sim 1500$ K.
The observed jet/counterjet brightness asymmetry is consistent
with an intrinsically symmetric outflow with extinction from
a dense, circumstellar structure of $\sim 6''$ size (along
the outflow axis), and with a mean visual extinction, A$_V$$\sim$ 11 mag.
\end{abstract}


\section{Introduction}

HH~1 and 2 were the first detected HH objects (Herbig 1951; Haro 1952).
Since then, they have played an important role in the study of outflows
from young stars, particularly because most of the general characteristics
of HH outflows were first seen in HH~1 and 2 (see the review of Raga et
al. 2011).

The first measurement of proper motions in HH objects (Herbig \& Jones
1981) showed that HH~1 and 2 formed part of a bipolar outflow.
The source of this bipolar outflow (centered between HH~1 and 2)
was discovered in the radio continuum by Pravdo et al. (1985). This
``VLA~1'' source was later shown to have a jet-like structure (of
$\sim 1''\times 0''.2$, see Rodr\'\i guez et al. 2000), aligned with
the HH~1/2 axis.

A jet-like structure directed towards HH1 (the ``HH1-jet'')
is observed at optical
wavelengths (Strom et al. 1985 point out this feature, which was
also visible in older images of the region). The base of this
jet-like structure approaches the position of the outflow source
(VLA~1) in images taken at progressively longer IR wavelengths
(Roth et al. 1989).

Reipurth et al. (2000) presented optical and IR images of the
HH1-jet region obtained with the HST. Their NICMOS H$_2$~2.12$\mu m$
and [Fe~II]~1.64$\mu m$ images show that the emission of the HH1-jet
extends to within $\sim 2''$ from the VLA~1 outflow source.
It is well known that all these multiwavelength jets are 
a manifestation of the same astrophysical phenomena
(see e.g. the review by Raga et al. 2010).
The fact that the observed HH1-jet emission does not extend
to the position of the VLA~1 source and that a counterjet (directed
towards HH~2) is not detected appears to be due to the presence of
a dense molecular structure approximately centered on VLA~1 (see
Torrelles et al. 1994; Choi \& Lee 1998 and Cernicharo et al. 2000).

In this paper we present new {\it Spitzer} observations in which
the counterjet (directed from the VLA~1 source towards HH~2) is
detected for the first time, and ending nearly thirty years
of speculation about the nature of its absence.

\section{Observations}

The observations of HH~1/2 system come from the GTO Spitzer Space Telescope
program by Giovani Fazio (PID 43) on the 'Orion Streamers' obtained with
the infrared camera IRAC (Fazio et al. 2004) in February 2004.
The data, consisting of the basic calibrated frames or BCDs,
have been recovered from the Spitzer Legacy Archive, version S8.18.
The original surveyed area is large and covers approximately 
0.75\arcdeg$\times$3.3\arcdeg in the four bands 
(1, 2, 3, 4) = (3.6, 4.5, 5.8, 8.0) \mum. The data were collected using the 
High-Dynamic-Range (HDR) mode with a 12sec integration time for the 'long' 
frames and 0.6sec for the 'short' ones. For this work we have used 
the 'long'  frames only, with a total integration time of 48sec per pixel.

The BCDs were then reprocessed with the HiREs deconvolution software AWAIC
(A WISE Astronomical Image Co-Adder) developed by the Wide Field Infrared 
Survey Explorer (WISE) for the creation of their Atlas images (see e.g.
Masci \& Fowler 2009).\footnote[3]{http://wise2.ipac.caltech.edu/staff/fmasci/awaicpub.html}
The AWAIC software optimizes the coaddition of individual frames by making
use of the Point Response Function (PRF) as an interpolation kernel, to avoid
flux losses in undersampled arrays like those of IRAC, and also
allows a resolution enhancement (HiRes) of the final image, by removing its
effect from the data in the deconvolution process.
A similar method has been applied on the {\it Spitzer} data of young stellar outflows
like HH 46/47 (Noriega-Crespo et al. 2004a; Velusamy et al. 2007) 
and Cep E (Moro-Mart\'\i n et al. 2001; Noriega-Crespo et al. 2004b; 
Velusamy et al. 2011), quite successfully. On IRAC images, the HiRes 
processing enhances the angular resolution from the standard
$\sim 2$\arcsec~to $\sim$ 0.6\arcsec~- 0.8\arcsec~(Velusamy et al. 2007).

 In Figures 1 and 2 we show the images of the HH~1/2 system obtained in the four
IRAC bands (3.6, 4.5, 5.8 and 8.0\mum) before (i.e. standard coaddition of the BCDs)
and after the HiRes reprocessing (iteration 25), respectively. At the iteration 25
the HiRes AWAIC processing reaches an optimal angular resolution, simultaneously
preserving most of the structure of the surrounding diffuse emission, for instance
that of the arc near the South of the images. Some small artifacts, however, are 
present in all bands, indicating that not further iterations are needed.
Figure 3 shows a closer look of the 4.5\mum~image, where the newly discovered
counterjet is the brightest, marking  as well some of the well known optical
knots of the HH~1 and 2 objects, and the position of the VLA~1 source.
The bright infrared source along the outflow symmetry axis, the so called
Cohen-Schwartz stars (C-S), that once was thought to be the driving source
(Cohen \& Schwartz 1979) is also indicated, plus the VLA~2 source that
drives the HH~144 outflow (Reipurth et al. 1993)

One can appreciate the performance of the HiRes processing and the ability of
the mid-IR observations to discover new features by comparing the processed IRAC 4.5\mum~
with archival data at optical and near-IR observations of the HH~1/2 system
obtained by the {\it Hubble Space Telescope}, with $\sim 5$ times better angular
resolution (Fig. 4). The optical image was taken by the Wide Field Planetary Camera 2 (WFPC2)
in 2007 (Hartigan et al. 2011) at [S~II] 6717/31~\AA, while the near-IR was taken 
with the NICMOS3 camera in 1998 at v=1-0 H$_2$ 2.12\mum~emission line (Reipurth et al.
2000). It is quite remakable how similar are the vibrational (at 2.12\mum) and rotational H$_2$ emission 
that arises from the S(9), S(10), S(11), S(12) transitions at
4.952, 4.408, 4.180 and 3.996\mum, covered by IRAC 4.5\mum~band.
About halfway between the VLA~1 source and HH~2, there are two or three
condensations detectable at 2.12 and 4.5\mum, and given their positions
could belong to counter flow as well.

\begin{figure}
\centerline{
\includegraphics[width=0.475\textwidth]{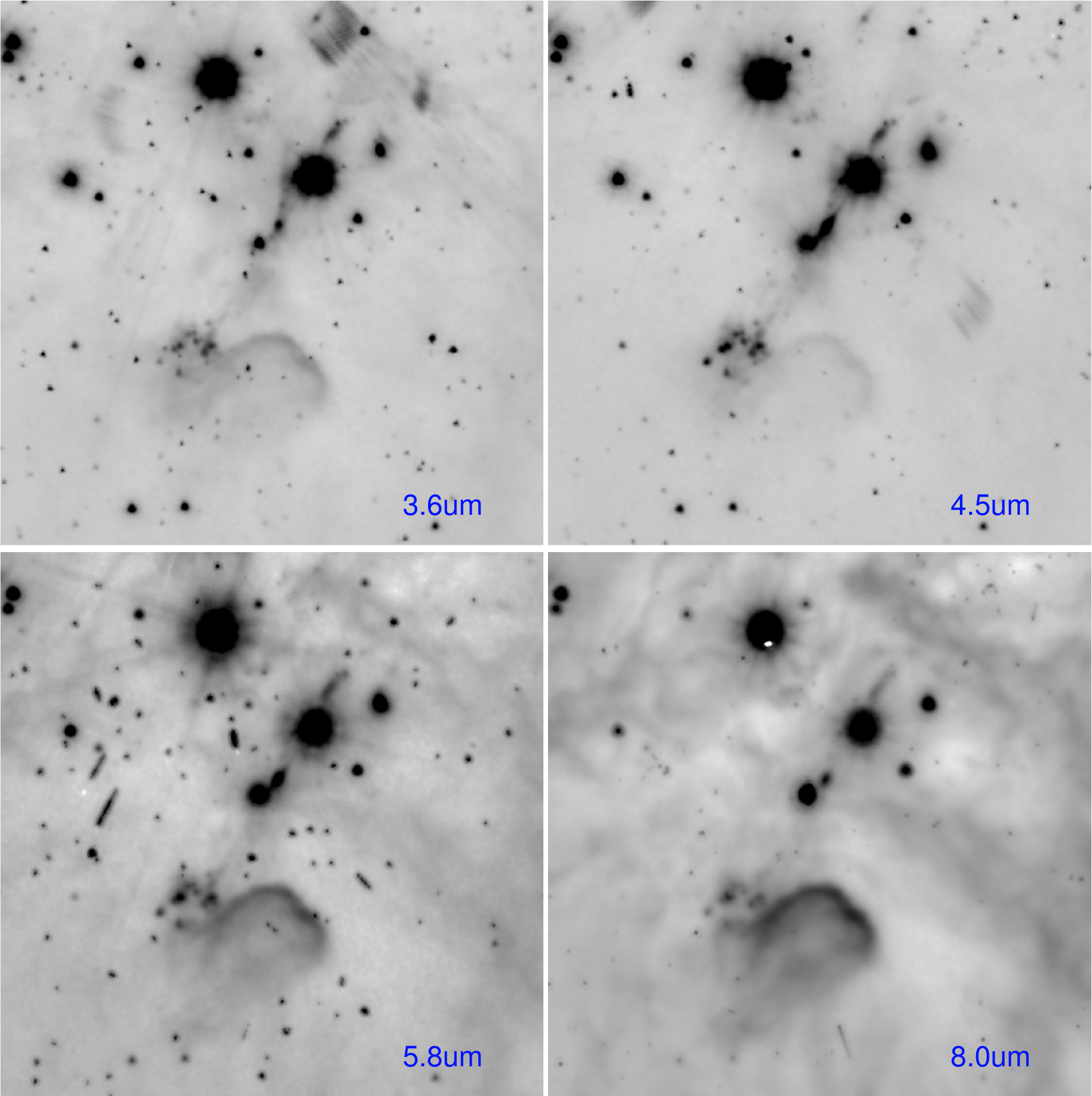}}
\caption{IRAC maps of the HH~1/2 system, from top left clockwise, at
3.6, 4.5, 5.8 and 8.0\mum; using a inverse grayscale where dark regions 
represent high intensity. The field-of-view is $\sim$ 5\arcmin. 
North is up and East is left.}
\label{fig1}
\end{figure}

\begin{figure}
\centerline{
\includegraphics[width=0.475\textwidth]{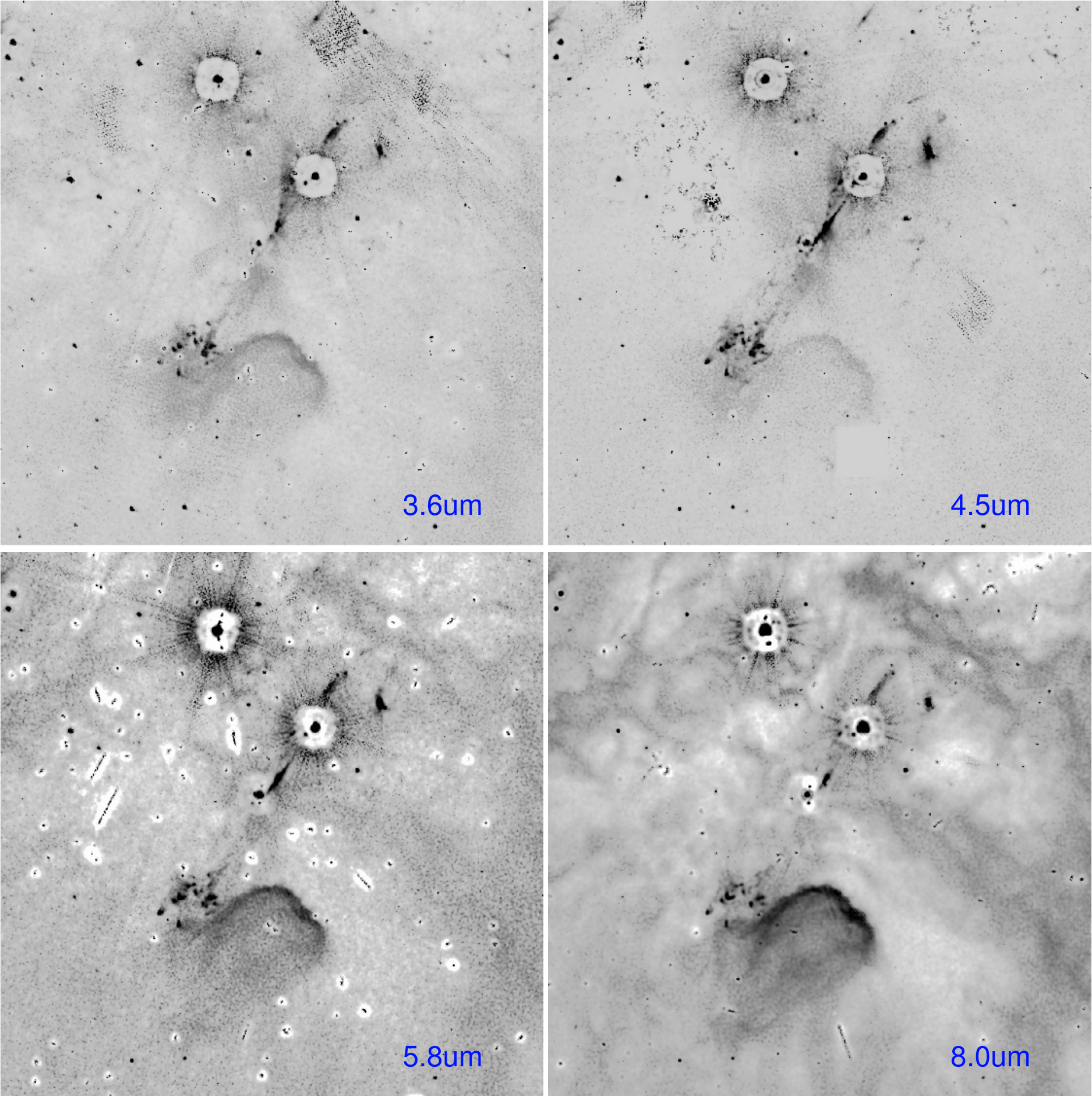}}
\caption{IRAC observations of the HH~1/2 system as in Fig 1,
after 25 iterations using the HiRes AWAIC algorithm.}
\label{fig2}
\end{figure}

\section{The HH~1 jet and counterjet}

A closer look of the [SII] 6717/31~\AA, 2.12, 4.5 and 8.0\mum~emission around
the jet/counter region is shown in Fig. 5. A detailed analysis of the optical
and near-IR HH~1-jet properties has been carried by Reipurth et al. (2000), where
they showed how the H$_2$ 2.12\mum~and [Fe~II] 1.64\mum~emission arise closer
to the VLA~1 source, only 2.5\arcsec~NW of it, than that of [S~II]. Their analysis
on the extinction, using the fact that the [Fe~II] 1.54\mum~and [S~II] 0.67\mum
have similar excitation energies and ionization potential, and that their ratio
is nearly constant for weak shocks, allowed them to determine a 4 mag increase
in the $\sim 5$\arcsec~that the nir-IR emission gets  closer to the VLA~1 source.

In the mid-IR at 4.5\mum~both the jet and counterjet are clearly detected,
and both can be traced back to the VLA~1 source. The jet/counterjet
path, unfortunately, is very close to two bright infrared sources; o
n the NE, the C-S source and at $\sim 10$\arcsec~SE of VLA~1 by a fainter one.
And so it is possible that very close to the circular edges were the AWAIC HiRes
algorithm suppresses the sources, that couple of knots are affected by
this artifact. Conservatively, the jet has a $\sim 12.9$\arcsec~length
in the NW direction arising from VLA~1, while the counterjet has
a $\sim 10$\arcsec~length in the opposite direction.
Both the optical and the mid-IR jet extend out
to the same NW position, the A$_{j}$ knot (Eisl\"offel, Mundt \& B\"ohm 1994).
Anf finally, at both 4.5 and 8.0\mum~couple of arcseconds NW of VLA~1, the emission broadens
following the structure observed at the base of the 2.12\mum~jet, the so called
'X' nebula (Reipurth et al. 2000). This nebula seems to be connected with the knots
East of the jet; they do have a similar structure at 2.12 and 4.5\mum. 

\begin{figure}

\centerline{
\includegraphics[width=0.475\textwidth]{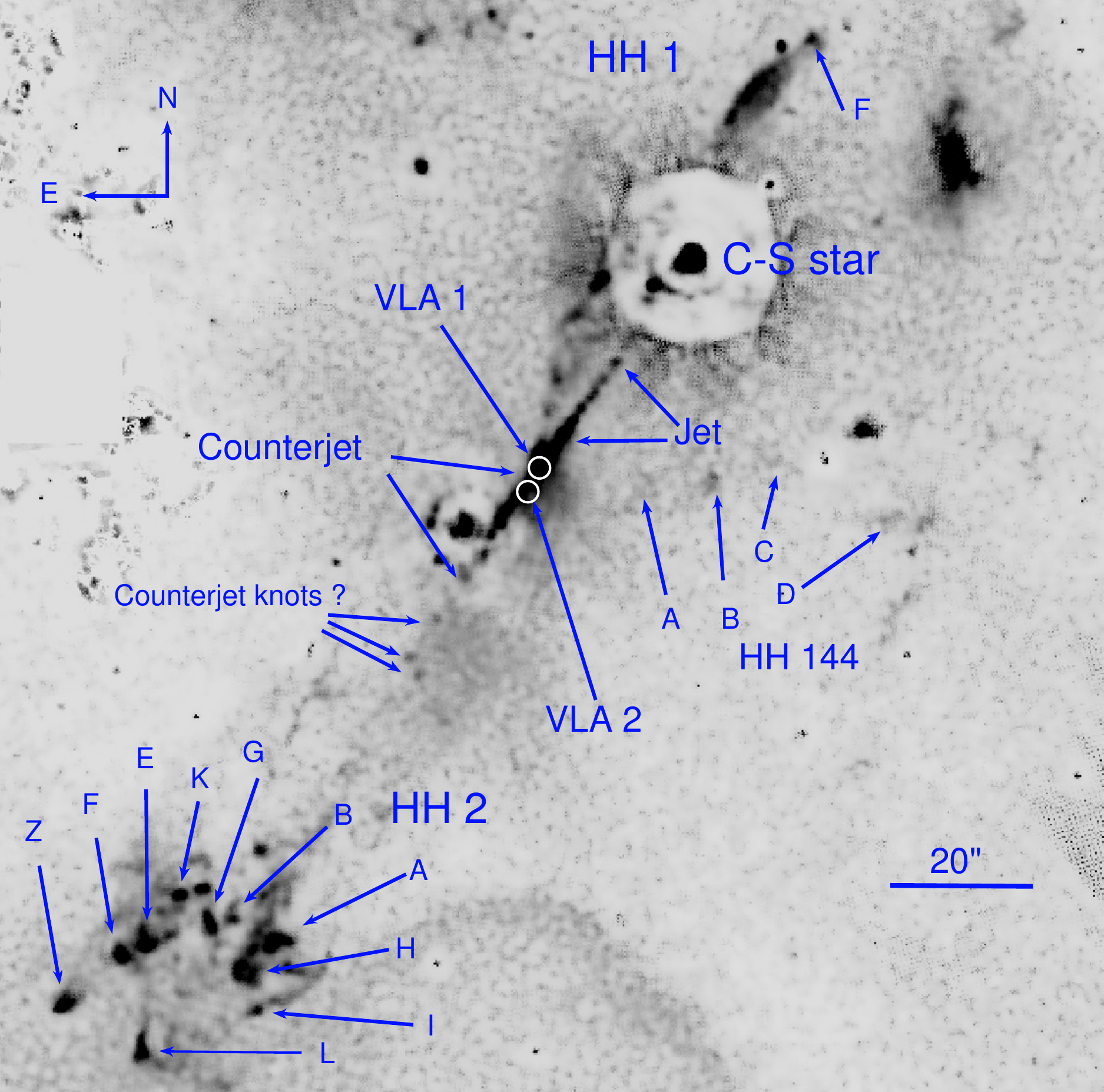}}
\caption{A closer view of the H 1/2 system at 4.5\mum, after the HiRes AWAIC processing,
where the jet and counterjet are best detected. The names
of some of the well known optical knots of HH~2 (SE) and HH~1 (NW) are included 
(see e.g. Raga, Barnes \& Mateo 1990), as well as those of HH~144 flow (Reipurth et al. 2000)
its driving source (VLA~2), and the bright IR Cohen-Schwartz (C-S) source 
(Cohen \& Schwartz 1979).}
\label{fig3}
\end{figure}

\subsection{The Medium Surrounding the Counterjet}

In Figure 6, we present 4.5 and 5.8\mum\ intensity tracings along
the jet/counterjet system. In order to obtain these tracings, we have defined 
an axis parallel to the direction of the outflow (the $x$-axis of Figure 6, with $x=0$
corresponding to the position of the VLA~1 source, and positive $x$ towards
the NW), and averaged the intensity in a direction perpendicular to the
outflow, in a box extending $\pm 3''$ to each side of the outflow axis. A
background was computed from contiguous boxes on each side of the outflow (of
$1''.5$ widths), and subtracted at all positions $x$ from the axial box in
order to obtain the jet emission.

\begin{figure}
\centerline{
\includegraphics[width=270pt,height=160pt,angle=0]{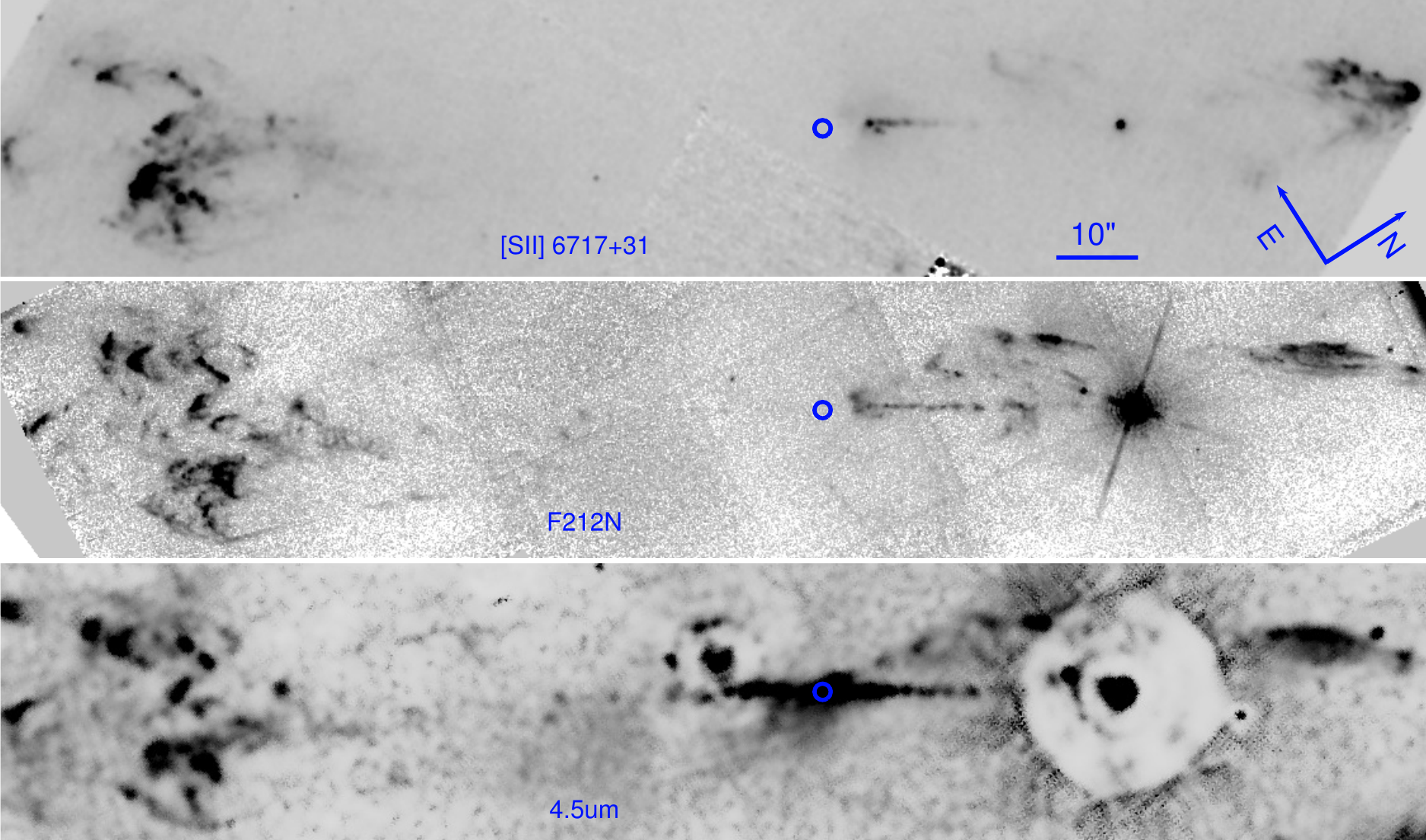}}
\caption{\label{fig4} {A comparison of the HH~1/2 system, from left to right  at optical
(WFPC2 [SII] 6717+31 \AA), near-IR (NICMOS 2.12\mum) and mid-IR wavelengths 
(IRAC 4.5\mum~HiRes [25 iterations] processing). 
The position of VLA~1 source is marked with a 3\arcsec~in diameter circle.}}
\end{figure}

\begin{figure}
\centerline{
\includegraphics[width=260pt,height=130pt,angle=0]{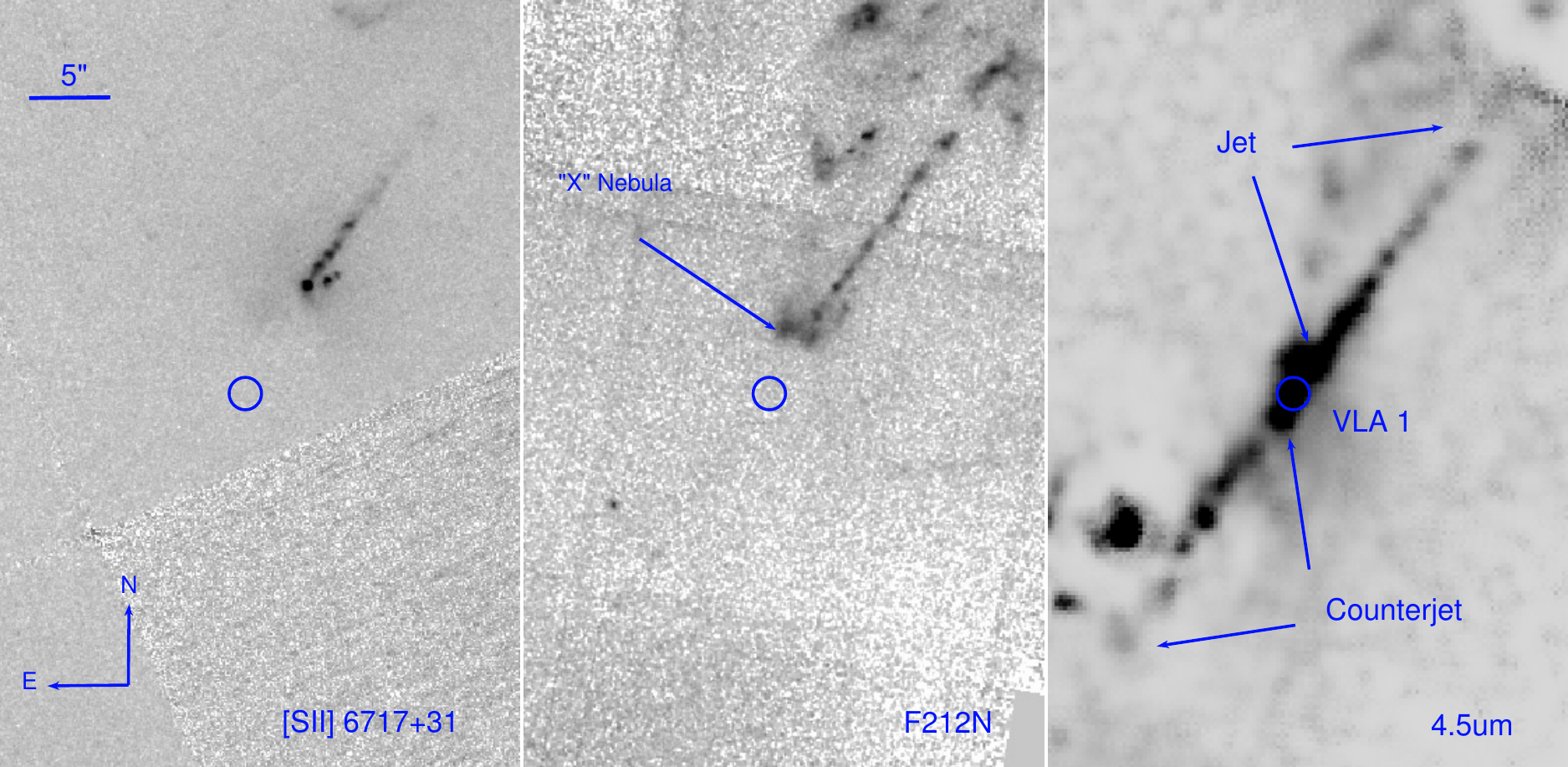}}
\caption{\label{fig5} {A closer view of the jet/counterjet region, using once again the 
{\it Hubble Space Telescope} observations in [SII] and 2.12\mum, and compared with
those at IRAC 4.5 HiRes (25 iterations) processing.
The position of VLA~1 source is marked with a 1\arcsec~in diameter circle.}}
\end{figure}

The intensity tracings show that the emission from the NW jet is generally
stronger than the counterjet emission. The bottom frame of Figure 6 shows
the jet/counterjet intensity ratios as a function of distance from the VLA~1
source at 4.5 (left; broken line) and 5.8 (right; solid line).
It is clear that the jet/counterjet intensity
ratio is $>1$ at 5.8\mum (at all positions).
At 4.5\mum, the jet/counterjet intensity
ratio is $>1$ out to $\sim 8''$ from the source, and $<1$
at $\sim 10''$ from VLA~1. This region (with jet/counterjet
intensity ratio $<1$) is associated with the last knot along
the counterjet, seen close to the stellar source $\sim 12''$
to the SE of VLA~1 (see Figures 3 and 5), and might therefore be
contaminated by the deconvolved point spread function of the star.
Therefore, we conclude that the observations are consistent
with jet/counterjet intensity ratios $>1$ at all positions
both at 4.5 and 5.8\mum\ (at least out to $\sim 8''$ from
the VLA~1 source, see Figure 6).
If one assumes that the jet and counterjet emission is
intrinsically symmetric, one would attribute the
position-dependence of the jet/counterjet intensity ratio
($I_j/I_{cj}$, see the bottom frame of Figure 6)
to the extinction produced by a high density clump
surrounding VLA~1. The fact that $I_j/I_{cj}$ reaches
a peak (with a value of $\approx 4.5$ at 4.5\mum,
and of $\approx 3$ at 5.8\mum, see Figure 6) at $x\sim 3''$
would then indicate that the high density clump has a
projected diameter of $\sim 6''$, and that for larger values
of $x$ the counterjet emerges from behind the clump (lowering
the observed $I_j/I_{cj}$ ratio). This $\sim 6''$ size is consistent with
the extension along the outflow axis of the flattened H$^{13}$CO$^+$ clump
observed by Choi \& Lee (1998), surrounding VLA~1.

The formation of stellar outflows arising from young stellar objects (YSOs) are expected
to be symmetric, since there is a preferential rotational axis and to the best of our understanding
of the formation of proto-stellar jets requires a magnetohydrodynamical coupling of the accreted gas
with the spherical symmetric protostar, likely to have also a symmetric bipolar magnetic 
structure (see e.g. Pudritz et al. 2007,  and references therein).
The accreted gas is provided by a disk-like structure created as a result of the original 
cloud spinning during the protostellar collapse (see e.g. Klein et al. 2007, and references therein).
One could imagine, however, that the accretion process does not need to be symmetric,
and could feed the protostar in a time dependent alternative way (e.g. one star pole at the time).
If this was the case, then the symmetry of the stellar jet could be broken, at least over a 
certain period of time, with no material ejected in one direction or another. And if this
was the case then one could trace back the steps to understand where (e.g. given the morphology
of the protostar's magnetic field, or its coupling with the accretion disk or the transfer
of gas and angular momentum, etc) this symmetry is broken.
A true asymmetric young stellar jet could have profound consequences in our understanding
of the low mass star formation process.  For nearly thirty years we have wondered if 
the HH~1/2 system had a counterjet, although as mentioned in the introduction
evidence for very dense gas (n(H$_2$ = $10^4$ \cc) structure around VLA~1,
using Ammonia as a tracer (Torrelles et al. 1994), suggested that
extinction was playing a major in hidding that component of the flow.

The NH$_3$ observations have an angular resolution
of $\sim 4$\arcsec~($\sim 2\times10^3$ AU at the distance of Orion), sampling
very well a 2\arcmin~$\times$3\arcmin~region covering the HH~1/2 outflow, 
and enough to distinguish a pancake-like structure perpendicular to the outflow axis
and a East-West temperature gradient in it that indicates a further asymmetry.
It is no a simple disk-like structure, but certainly a dense structure consistent
with the idea of "a collapsing interstellar ($\sim$ 0.4pc) toroid around
VLA~1" (Torrelles et al. 1994).
NH$_3$ has been used as tracer of high density and low temperature molecular 
gas around young stellar outflows for more than two decades. The NH3(1,1) transition requires
H2 densities higher than 5$\times 10^3$\cc~to be detected reliably (Torrelles et al. 1983). 
Typical values around YSOs range from n(H2) = 5$\times 10^3$\cc~to $\sim 10^5$\cc~
and Temperatures of 15 - 30 K (Torrelles et al. 1983).
The structure around the VLA~1 in HH~1/2  has n(H2) $\sim 10^4$ \cc and 12 K.
Similar conclusion ws reached by Choi \& Zhou (1997) using other high density molecular tracer (HCO$^+$). 
Slightly better angular resolution observations (4.3\arcsec$\times$2.8\arcsec~beam, Choi \& Lee 1998)
have traced even closer to the VLA~1 source the toroidal structure.
Very close to the VLA~1 source, within a diameter of 4 AU, Cernicharo et al. 2000 estimated a
very high visual extinction (80-100 magnitudes), but they were let's concerned with the 
"toroidal structure". In Figure 7 we show a comparison of the NH$_3$(1,1) observations
by Torrelles et al. (1994) with the mid-IR emission at 4.5\mum. Two things are worth noticing,
First, that high density molecular gas does indeed prevent us from detecting the counterjet
at shorter wavelengths, and second, that the morphology of this dense structure is likely
to prevent us to detect the HH~144 counterflow.

Independently of detailed structure of the "toroid", it is clear that a large column density
of material is a long the line-of-sight of the counterjet, and this is higher than in 
line-of-sight of the visible jet.  An estimate of the extinction around the counterjet is possible, 
although requires some necessary assumptions.
For instance, based on symmetry one could assume that the emission properties at 3.6 and 4.5/mum~ 
of jet and counterjet are very similar over the same area, and that any difference between the jet (North) 
and the counterjet (South) are due only to extinction. A comparison of the ratio of the surface 
brightness of the  3.6 to 4.5/mum~images, indeed shows smaller values on the South with respect 
to the North. The average ratio of the  North and South sections of the jet, 
using a 2.4\arcsec$\times$39.0\arcsec~box, give values of 0.40$\pm$0.27 and 0.20$\pm$0.16, respectively, 
i.e. the counterjet is twice as faint when compared to the jet. Two recent studies have analyzed the properties 
of the mid-IR extinction (Indebetouw et al. 2005, Cambresy et al. 2011),
in regions located in the Galactic plane. i.e. RCW49 and the Trifid, with very similar 
results, so one can assume that such extinction  approximately holds in the environment around the HH~1/2 system. 
Taking the ratios of extinction at Visual, K$_s$ (2.2\mum), 3.5 and 4.5\mum~to be 
A$_K$/A$_V$ = 0.112, A$_{3.6}$/A$_K$ = 0.611 and A$_{4.5}$/A$_K$ = 0.500 (Cambresy et al. 2011), 
and neglecting the extinction at 4.5/mum~(since both jet/counterjet are detected) then at 3.6\mum~
at the counterjet one obtains, A$_{3.6}$ = 0.75, that corresponds to A$_K$ = 1.2 and 
A$_V$ = 11.0 magnitudes, respectively.

\begin{figure}
\centerline{
\includegraphics[width=0.475\textwidth]{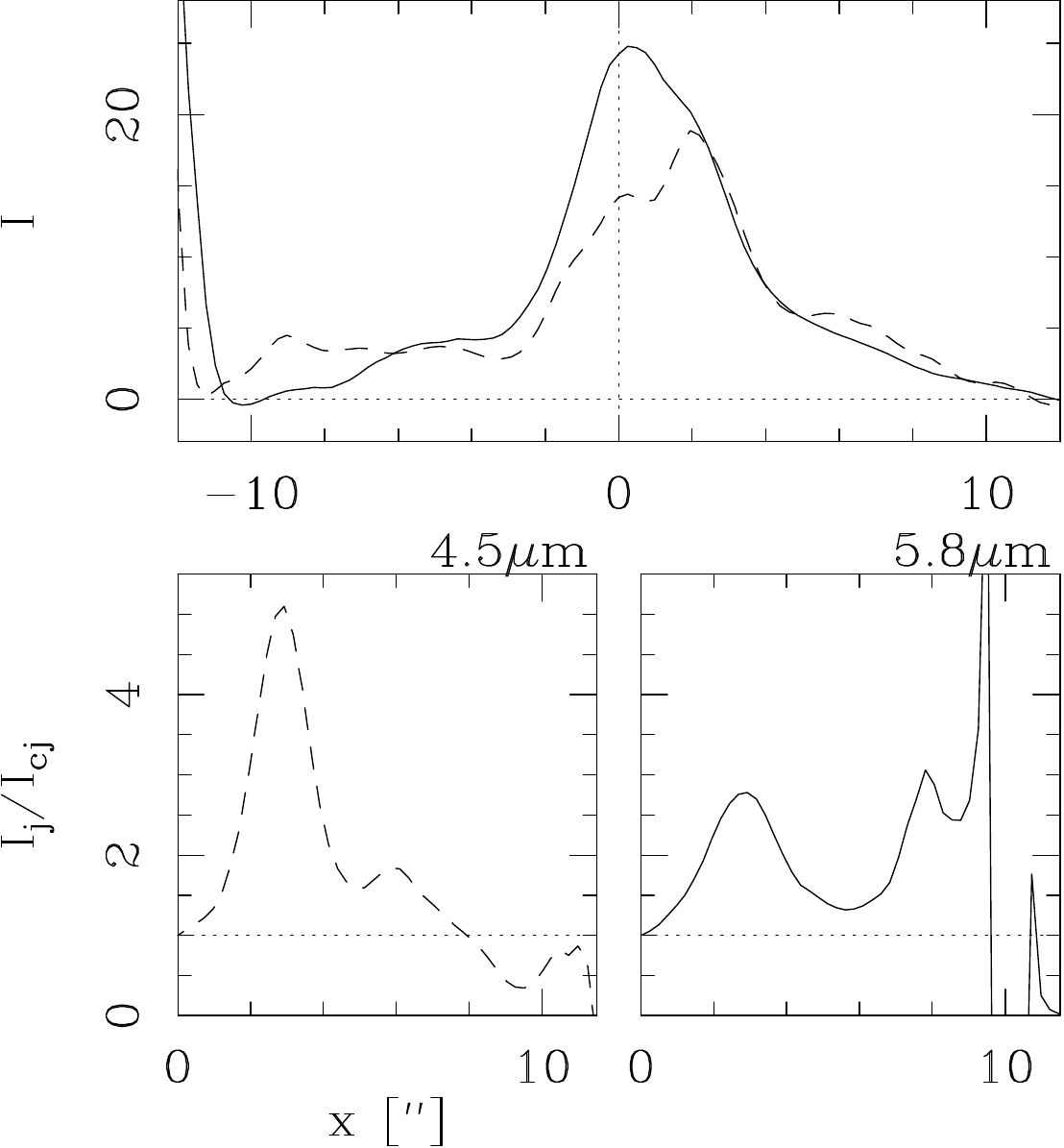}}
\caption{Top frame: surface brightness (averaged across the width of the jet,
in MJy/sr) at 4.5 (dashed line) and 5.8\mum~(solid line)
along the NW jet (positive $x$) and SE counterjet (negative $x$) centered on
the VLA~1 source, showing the asymmetry
of the two outflow lobes. Bottom frame: the jet/counterjet intensity ratio as
a function of distance from the VLA~1 source at 4.5 (left panel; broken line) and
5.8 \mum~(right panel; solid line).}
\label{fig6}
\end{figure}

\begin{figure}
\centerline{
\includegraphics[width=0.475\textwidth]{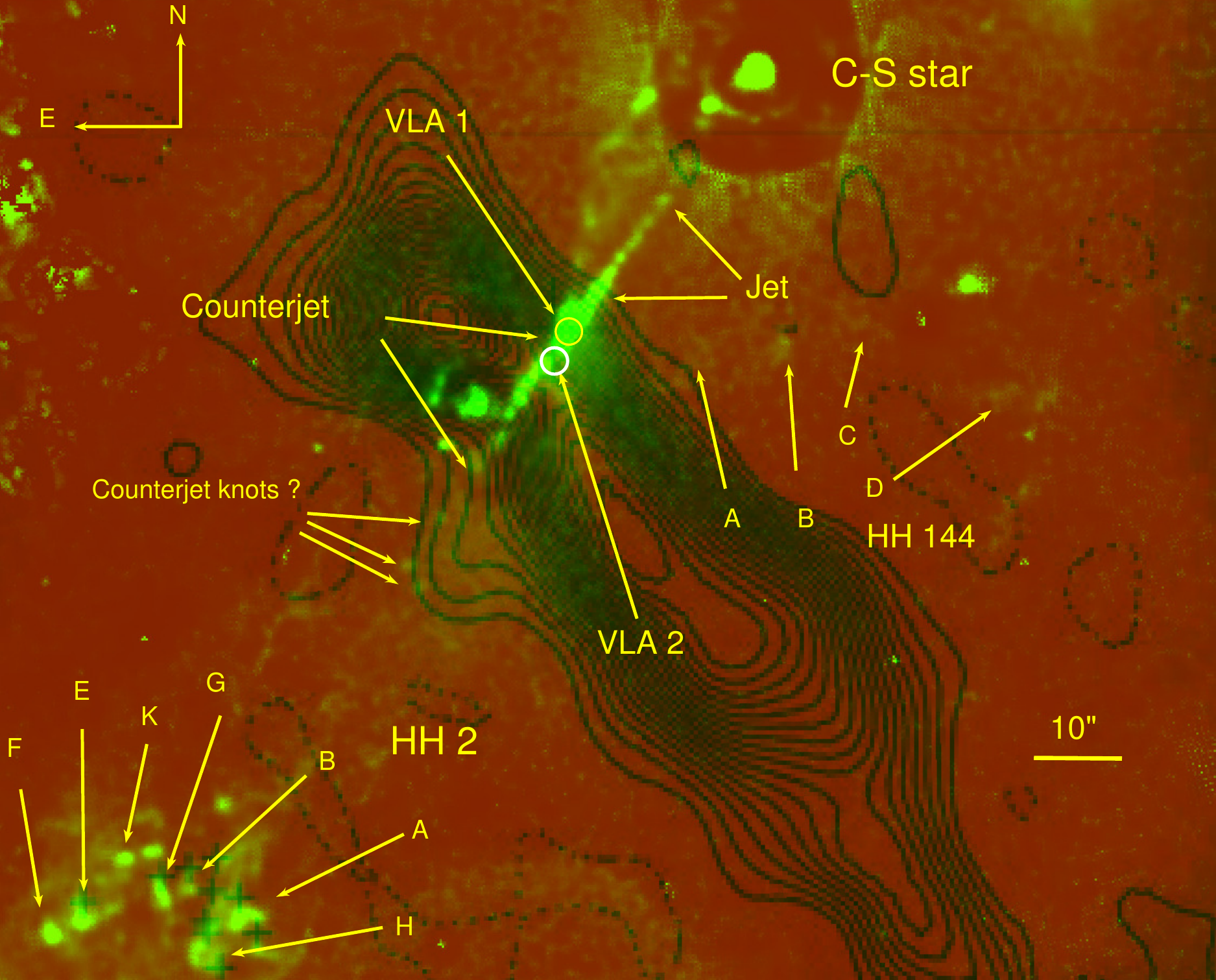}}
\caption{A comparison of the dense structure detected in NH$_3$(1,1) 
at 10.5 \kms~(dark contours; from Torrelles et al. 1994)
with the emission at 4.5\mum~around the HH~1/2 jet/counterjet region.}
\label{fig7}
\end{figure}

\section{The HH~1/2 mid-IR knots}

Since some of the first IRAC images of stellar outflows from {\it Spitzer}
(e.g. Noriega-Crespo et al. 2004a,b, Morris et al. 2004), it has been clear
that the bands at 4.5 and 5.8\mum\ are particularly suitable to study them. 
There are seven pure rotational H$_2$ emission lines, from S(12) at 3.996\mum~to S(6) 6.107\mum  
that fall within their bandpasses. The 8\mum~band includes some
bright v=0-0 H$_2$ lines (e.g. S(5) 6.907\mum), but the emission in this band tends to be
dominated by broad band features from very small UV stochastically heated 
dust particles or Polycyclic Aromatic Hydrocarbons (PAHs) (see e.g. Tielens 2008).
Recently, Ybarra \& Lada (2009) have used a combination of IRAC colors to uniquely identify
the thermal emission arising from collisionally excited H$_2$ as a function of the
HI gas density and temperature, and thus provide a reliable way to study the thermal
structure of the stellar outflows by using the ([3.6]$-$[4.5]) and ([4.5]$-$[5.8]) colors.

Some of the knots of the HH~1/2 system are clearly compact
and at a first approximation one can treat them as point sources
and transform their photometric fluxes into magnitudes 
(see e.g. ``IRAC Instrument Handbook''
\footnote[4]{http://irsa.ipac.caltech.edu/data/SPITZER/docs/irac/iracinstrumenthandbook}
,  \S 4.11.1, 'Best Practices for Extended Sources'). One needs to be aware
that interpreting surface brightness measurements into colors
can be off by 5\% - 10\% (``IRAC Instrument Handbook'', \S 4.11.1).
We adopt a 10\% uncertainty across bands in our measurements,
and use the standard IRAC zero points (``IRAC Instrument Handbook'',
Table 4.1) of F$_{\nu0}$ = 280.9, 179.7, 115.0 and 64.9 Jy for
the 3.6. 4.5, 5.8 and 8.0\mum~bands, respectively.
The results are presented in Table 1 and display in a color-color
diagram in Figure 8, following Ybarra \& Lada (2009).
The diagram shows the properties of shocked excited H$_2$
as a function of constant HI density (dotted line) and temperature (solid line).
At a temperature higher than 4000 K, the H$_2$ is likely to
be dissociated (Ybarra \& Lada 2009).

The excitation properties of the HH~1/2 knots at optical and UV
wavelengths are very well known (see e.g. Solf, B\"ohm \& Raga 1998;
Solf et al. 1991; B\"ohm \& Solf 1992; Schwartz et al. 1993;
B\"ohm, Noriega-Crespo \& Solf 1993; Moro-Martin et al. 1996;
Molinari \& Noriega-Crespo 2002)
and are the result of their shock velocity and relative geometry. 
In general, high excitation knots are compact and interact through strong
shocks (60 - 100 \kms) with the surrounding medium, while
low excitation emission arises from weaker shocks or the
``wings'' of bowshock-like condensations. For example, knots
like HH~1F, HH~2A, HH~2G and HH~2H are high excitation,
while HH~2I, HH~2L are low excitation (Raga, B\"ohm, \& Cant\'o 1996).
 For the time variable jet, where the relative velocity between
knots determines essentially the shock velocity, the overall excitation
is known to be low and the shocks to be weak (Reipurth et al. 2000).
Stronger shocks will produce higher post-shock 
gas densities and temperatures, but could also dissociate
H$_2$; stronger H$_2$ emission indeed is expected
to arise from low velocity shocks (either J-type or C-type
(see e.g. Draine \& McKee 1993, for a review).

\begin{figure}
%\epsscale{1.0}
%\plotone{f8.eps}
\centerline{\includegraphics[width=0.475\textwidth]{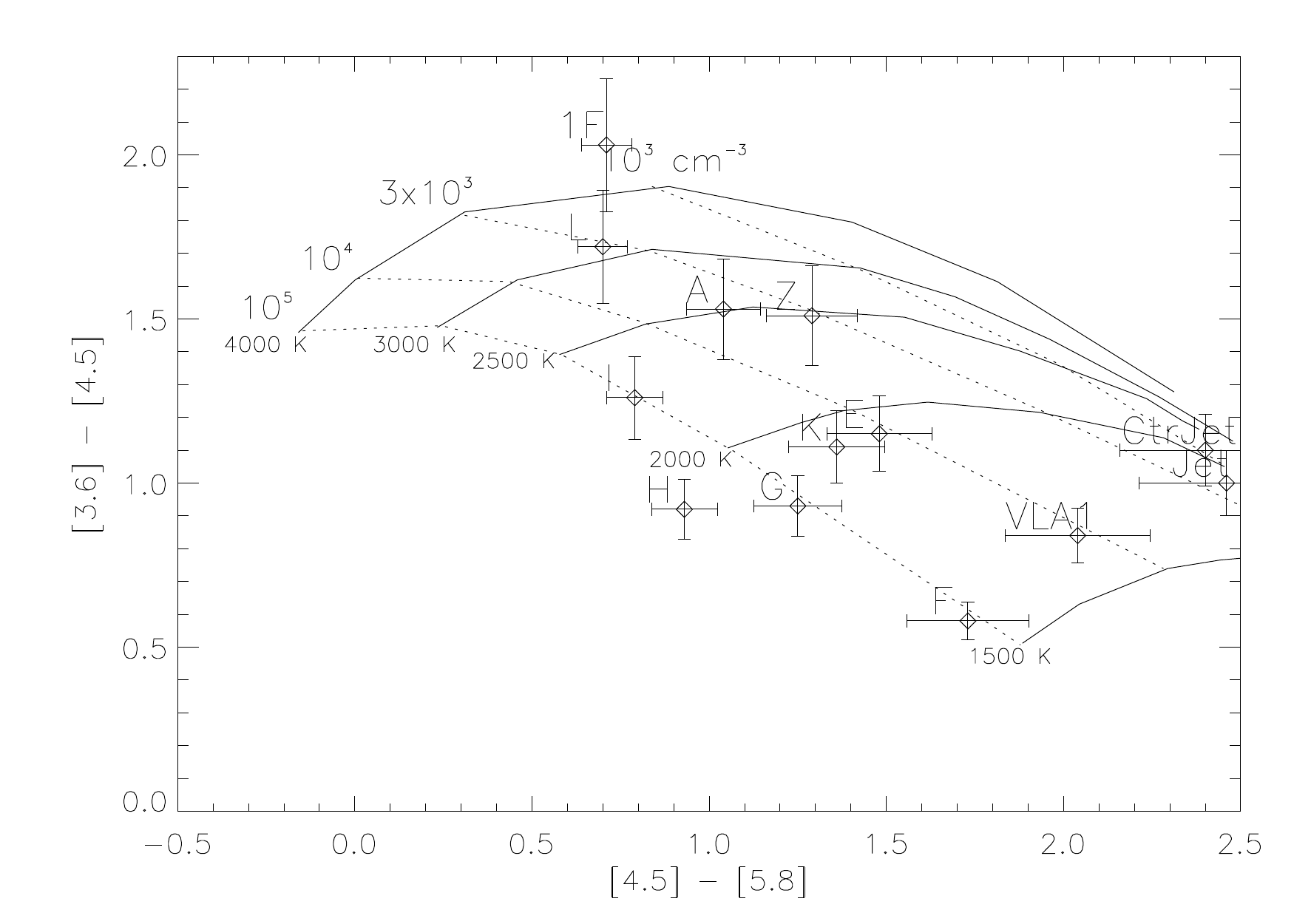}}
\caption{IRAC color-color diagram (after Ybarra \& Lada 2009) 
of compact knots in the HH 1/2 outflow with optical counterparts, 
including HH~1F (1F), HH~2 (A, E, F, G, H, I, K, and Z) and 
regions around the jet (Jet), counterjet (CtrJet) and the VLA~1 source (see Table 1).}
\label{fig8}
\end{figure}

So how the properties of the mid-IR knots, dominated
by emission of H$_2$ rotational lines at 4.5 and 5.8\mum,
compare with those of the optical knots? One of the highest
optical excitation knots, HH~1F, is slightly above the 4000 K
model, the region of the diagram where indeed H$_2$ dissociates.
The jet and  counterjet, low excitation regions,
do fall within the low temperature $\sim 1500$K and low density
($\sim 10^3$ \cc) models. Notice also that their IRAC colors
are very close to each other, providing support to the
assumption that their emission properties are nearly identical.
The low excitation HH~2F and the region around VLA~1,
are closer to the low 1500 K temperature model, but at higher gas density
$\sim 10^5$ and $10^4$ \cc, respectively.
The rest of the mid-IR knots are found between these two extremes.
We notice that HH~2H and G, that are along the highest density
model (10$^5$\cc) are also colder than the low excitation knots
HH~2I and L, and this suggests that faster cooling at higher densities could
play a role in reducing the excitation in the mid-IR.

\section{Summary and Conclusions}

We present previously unpublished {\it Spitzer} observations of
the HH~1/2 outflow. The outflow is detected in the four
IRAC channels (3.6, 4.5, 5.8 and 8.0\mum). In the 4.5\mum\
channel, a well collimated jet/counterjet system is seen
emanating from the VLA~1 radio source, which represents the
discovery of the SE counterjet (previous optical and IR
observations only detected the NW jet directed towards HH1.

We find that the ratio between the jet and counterjet
emission is strongly dependent on distance from the VLA~1 source.
The jet/counterjet ratio is $\sim 2$ times larger for
the 4.5\mum\ than for the 5.8\mum\ emission. This result
is consistent an extinction effect (which would be
stronger at shorter wavelengths). The fact that the jet/counterjet
ratio peaks at a distance of $\sim 3''$ from the VLA~1 source indicates that
the extinction is probably produced by a dense structure having
an extent of $\sim 6''$ along the outflow axis. Such a
dense molecular structure around VLA~1 has been observed by
Choi \& Lee (1998) and Cernicharo et al. (2000).
We also show that the large structure of the dense gas traced by NH$_3$.
plays role in hidding both the HH~1/2 and HH~144 counterjets.
Under the assumption that the ratio of 3.6 to 4.5\mum~is constant
for the jet/counterjet, we estimate a A$_V$ = 11 mag.

Finally, by using the IRAC colors ([3.6] - [4.5]) and ([4.5] - [5.8]) 
and assuming that outflow emission is dominated by shocked excited
H$_2$ (Ybarra \& lada 2009), we show that some of the compact mid-IR knots 
share similar excitation properties as those of determined for the optical knots
(Raga, B\"ohm \& Cant\'o 1996).

\acknowledgements
We thank Frank Masci for the development of AWAIC HiRes software
and making it available to us. We also thank Dr. Sean Carey for useful conversations,
and the anonymous referee for her/his careful reading of the manuscript.
This work is based in part on observations made with the {\it Spitzer Space 
Telescope} which is operated by the Jet Propulsion Laboratory, California 
Institute of Technology under NASA contract 1407.
And also based on observations made with the NASA/ESA Hubble Space Telescope, 
and obtained from the Hubble Legacy Archive, which is a collaboration between 
the Space Telescope Science Institute (STScI/NASA), the Space Telescope European 
Coordinating Facility (ST-ECF/ESA) and the Canadian Astronomy Data Centre (CADC/NRC/CSA).
The work of AR was supported by the CONACyT grants 61547,
101356 and 101975.

%%%\clearpage

\setlength{\bibhang}{2.0em}

%\clearpage

\begin{deluxetable}{rccccc}
\tablecaption{HH~1/2 System IRAC Colors\label{tbl:colors}}
\tablewidth{0pt}
\tablecolumns{6}
\tablehead{
\colhead{Knot} & \colhead{[3.6] - [4.5]}  & \colhead{[4.5] - [5.8]}
& \colhead{RA} & \colhead{Dec}  & \colhead{AperRad(\arcsec)}}
\startdata
  Z\tablenotemark{a}& 1.29$\pm$0.13 & 1.51$\pm$0.16 & 5h36m27.40s & -6d47m22.2s & 3.0 \\
  F & 1.73$\pm$  0.18&0.58$\pm$  0.06  &  5h36m26.84s & -6d47m15.0s & 1.8 \\
  L & 0.70$\pm$  0.07&1.72$\pm$  0.17  &  5h36m26.65s & -6d47m28.3s & 3.0 \\
  E & 1.48$\pm$  0.15&1.15$\pm$  0.12  &  5h36m26.58s & -6d47m13.2s & 1.8 \\
  K & 1.36$\pm$  0.14&1.11$\pm$  0.11  &  5h36m26.24s & -6d47m06.1s & 1.8 \\
  G & 1.25$\pm$  0.13&0.93$\pm$  0.10  &  5h36m26.00s & -6d47m10.1s & 1.8 \\
  A & 1.04$\pm$  0.10&1.53$\pm$  0.15  &  5h36m25.34s & -6d47m11.9s & 2.4 \\
  H & 0.93$\pm$  0.10&0.92$\pm$  0.10  &  5h36m25.65s & -6d47m17.2s & 2.4 \\
  I & 0.79$\pm$  0.08&1.26$\pm$  0.13  &  5h36m25.53s & -6d47m22.5s & 1.8 \\
 Jet& 2.46$\pm$  0.25&1.00$\pm$  0.11  &  5h36m22.44s & -6d45m58.1s &1.8 \\
CtrJet&2.40$\pm$  0.24&1.10$\pm$ 0.10  &  5h36m23.00s & -6d46m08.0s & 1.8 \\
 VLA~1&2.04$\pm$  0.20&0.84$\pm$  0.09  &  5h36m22.83s & -6d46m05.0s & 1.8 \\
 1F & 0.71$\pm$  0.07&2.03$\pm$  0.20  &  5h36m20.23s & -6d45m05.6s & 2.4 \\
\enddata
{\baselineskip=0pt
\tablenotetext{a}{Knot 8 at 2.12$\mu$m~(Davis, Eisl\"offel \& Ray 1994)}}
\end{deluxetable}

%%%%%\end{document}

\end{document}